\documentclass[twocolumn,           % Format : preprint, twocolumn
               showpacs,            % Pacs : showpacs, noshowpacs
               preprintnumbers,     % Preprint: preprintnumbers,
               aps,                 % Society: ...
               prd,          	    % Journal Style : pra, prb, prc, prd, pre,
               a4paper,             % Size : a4paper, ...
               groupedaddress,      % Affiliation (Title) : groupedaddress,
               nofootinbib,         % Footnote: footinbib, nofootinbib
               tightenlines,        % Remove additional spaces in a line
               floats,floatfix      % Floating pictures and tables
               ]{revtex4}

\usepackage{graphicx}
\usepackage{latexsym}
\usepackage{amsmath,amssymb}        % amssymb includes amsfonts
\usepackage[draft=false]{hyperref}

\begin{document}

%%--- DRAFTCOPY --------------------------------
%% Prints a large "DRAFT" diagonally across each page
%% Does not show up in TeXview
%% \typeout{Prints "DRAFT" on each page; does not show in TeXView}
%%------------------------------------------------

%======================================%
%<<<<<<<<<<<< TITLE PAGE >>>>>>>>>>>>>>%
%======================================%

\title{Newtonian Collapse of Scalar Field Dark Matter}
\author{F. Siddhartha Guzm\'an$^{1}$ and L. Arturo Ure\~{n}a-L\'{o}pez$^{2,3}$}
\affiliation{
$^{1}$Max Planck Institut f\"ur Gravitationsphysik, Albert Einstein Institut, Am M\"uehlenberg 1, 14476 Golm, Germany. \\
$^{2}$Astronomy Centre, University of Sussex, Brighton BN1 9QJ, United Kingdom. \\
$^{3}$Instituto de F\'{\i}sica de la Universidad de Guanajuato, A.P. E-143, C.P. 37150, Le\'{o}n, Guanajuato, M\'{e}xico.}
\date{\today}
\preprint{astro-ph/xxxx}

%======================================%
%<<<<<<<<<<<<< ABSTRACT >>>>>>>>>>>>>>>%
%======================================%

\begin{abstract}
In this letter, we develop a Newtonian approach to the collapse of galaxy fluctuations of scalar field dark matter under initial conditions inferred from simple assumptions. The full relativistic system, the so called Einstein-Klein-Gordon, is reduced to the Schr\"odinger-Newton one in the weak field limit. The scaling symmetries of the SN equations are exploited to track the non-linear collapse of single scalar matter fluctuations. The results can be applied to both real and complex scalar fields.  
\end{abstract}

\maketitle

%======================================%
%<<<<<<<<<<<<<< ARTICLE >>>>>>>>>>>>>>>%
%======================================%

The most recent cosmological observations \cite{max} support the Cold Dark Matter (CDM) as the standard model of cosmological structure formation. So far, the CDM model agrees reasonably well with many observations at large and galactic scales \cite{primack,prada}, though its predictions at subgalactic scales are still a matter of intense debate \cite{dekel}. However, the nature of CDM remains one of the most intriguing problems in modern cosmology. It is amazing that CDM can be so predictive and survive the confrontation with cosmological data just by assuming general features of dark matter such like, for instance, that it is made of weakly interacting massive particles (WIMPS). On the other hand, with CDM becoming a reliable paradigm of structure formation and its nature still uncertain, it is possible to test models of dark matter as alternatives to the WIMP hypothesis; models in which the physical processes can be much more under control.

One of such alternative models is the scalar field dark matter (SFDM), in which dark matter should be comprised of ultra-light scalar field particles \cite{lee,sin,arbey,peebles,miguel,fuzzy,mielke}. On the one hand, the cosmological evolution of SFDM and its linear fluctuations can match those of CDM. On the other hand, self-gravitating scalar configurations can reproduce some general properties of galaxy halos as observed today. Going further, it is possible to study the non-linear evolution of SFDM fluctuations to form gravitationally bound objects, as it was first shown in \cite{miguel}. This is done by numerically evolving the coupled Einstein-Klein-Gordon (EKG) equations for scalar field configurations \cite{miguel,seidel90,seidel91,miguel1}. 

In this letter, our main aim is to study the collapse of scalar field dark matter under realistic conditions. In this sense, we will try to complete the picture of structure formation with scalar field dark matter, from the growing of primordial fluctuations up to its non-linear collapse. Also, our intention is to draw a guideline as simple as possible, for which we will only take into account the simplest assumptions; a more detailed picture will be presented elsewhere.

First of all, we consider that SFDM is comprised of a minimally coupled real scalar field $\Phi$ endowed with a scalar field potential $V(\Phi)$, which accounts for the self-interactions of the scalar field. Even though the general form of the scalar potential can be complicated, we are interested only in those potentials that behave as $V(\Phi)=(m^2/2) \Phi^2$ at late times \cite{lee,sin,arbey,peebles,miguel}. Such a quadratic scalar potential is preferred because, in a homogeneous and isotropic Universe, the scalar field oscillates coherently around the minimum of the potential and then the scalar energy density $\rho_\Phi \simeq m^2 \Phi^2$ scales much the same as CDM. Moreover, the scalar field fluctuations in the linear regime also grow in the same manner as CDM fluctuations. Most of the works suggest that a realistic model of structure formation should consider a very light mass for the boson particles \cite{lee,sin,arbey,peebles,miguel,fuzzy}. We will use here the value $m=10^{-23} {\rm eV} = (2.09 \, \rm yr)^{-1} = (0.64 \, \rm pc)^{-1}$ (we are taking units such that $c=\hbar=1$; also $G=m^2_{\rm Pl}$ with $m_{\rm Pl}$ is the Planck mass), which is also a natural scale for time and distance within SFDM.

Due to the similarities of scalar field dark matter with the CDM model, it is reasonable to take as a guideline for the non-linear regime of structure formation the standard spherical collapse model \cite{liddle}. According to this, a spherical overdense fluctuation slowly separates out from the cosmological expansion, reaches a maximum expansion and then collapses under the influence of its own gravity only. The important instant here is that of maximum expansion, which is called the time of {\it turnaround}. 

Before turnaround, we do not expect violent processes occurring to the matter fluctuation, and that is why we consider the spherical collapse model as a good approximation for scalar fields. However, we indeed expect the gravitational collapse of the matter fluctuation to be determined by the intrinsic nature of the scalar fields. Hence, the conditions at turnaround will be our {\it initial} conditions for the gravitational collapse of scalar field dark matter.

According to the spherical collapse model, the non-linear density contrast at turnaround is $\delta^{\rm turn}_\Phi \equiv [\delta\rho_\Phi /\rho_\Phi]^{(\rm turn)}= 4.55$. As we said before, the homogeneous scalar energy density evolves as the standard cold dark matter does, and then the local value of the scalar energy density at turnaround is approximately given by
\begin{equation}
8 \pi G \, \Phi^2_{\rm turn} \simeq 13.55 \,\Omega_{0, \rm CDM} \left( 1+z_{\rm turn}\right)^3 H^2_0/m^2 \, , \label{e3}
\end{equation}
where $\Omega_{0, \rm CDM} \simeq 0.25$, $H_0=70 \, {\rm km \, s^{-1} 
Mpc^{-1}}$ is the current value of the Hubble parameter, and $z_{\rm turn} \sim {\rm few}$ is the redshift at turnaround. Thus, the local value of the scalar field at turnaround is quite small, of the order $ \sqrt{\kappa_0} \Phi^{\rm (local)}_{\rm turn} \sim 5.06 \times 10^{-10} $.

With this in mind, we proceed now to describe the gravitational collapse of a fluctuation. As it has been shown in \cite{miguel,seidel90,seidel91,miguel1}, the problem is well defined within General Relativity, of which we give here a brief description. Assuming spherical symmetry, the metric is written in the form
\begin{equation}
ds^2 = -\alpha^2 dt^2 + a^2 dr^2 + r^2 \left( d\theta^2 + \sin^2 \theta d\varphi^2 \right) \, ,\label{e5}
\end{equation}
where $\alpha(t,r),a(t,r)$ are functions determined by the matter distribution through the coupled Einstein-Klein-Gordon (EKG) equations,
\begin{eqnarray}
G_{\mu \nu} = 8\pi G \, T_{\mu \nu} (\Phi) \, , \, && \, \, \Box \Phi = m^2 \Phi \, . \label{e6}
\end{eqnarray}
Here, $G_{\mu \nu}$ is the Einstein tensor corresponding to the metric (\ref{e5}), and $T_{\mu \nu}(\Phi)$ is the scalar energy-momentum tensor. The KG equation arises from the Bianchi identities, and $\Box \Phi = \frac{1}{\sqrt{-g}}\partial_{\mu}[\sqrt{-g} g^{\mu \nu}\partial_{\nu} \Phi]$.

In principle, we should just numerically evolve Eqs. (\ref{e6}), but as we have to deal with weak gravity we find more appropriate to evolve the weak field limit of such system of equations, which by the way provides important technical simplifications that let us to have more physical insight. In our case, that limit arises when $\alpha^2-1,a^2-1,\sqrt{8 \pi G} \, \Phi < 10^{-3}$ \cite{seidel90,luis}.

As it was shown in \cite{seidel90}, such limit is found for complex scalar fields through a standard post-Newtonian treatment, and we now show how a similar procedure can also be applied to real scalar fields. We begin by writing such real scalar field and the metric coefficients in terms of the Newtonian fields $\psi,U,U_2,A,A_2$ as
\begin{eqnarray}
\sqrt{8\pi G} \, \Phi &=& e^{-i\tau} \psi(\tau, x) + {\rm C.C.} \, , \\
\alpha^2 &=& 1+2U(\tau,x) + e^{-2i\tau} U_2(\tau, x) + {\rm C.C.} \, , \\
a^2 &=& 1+2A(\tau,x) + e^{-2i\tau} A_2(\tau, x) + {\rm C.C.} \, ,
\end{eqnarray}
where we have also introduced the dimensionless quantities $\tau = mt, ~ x=mr$. Notice that only $U,A$ are real fields. Next, we assume that all the new fields are of order ${\cal O}(\epsilon^2) \ll 1$, and that their derivatives obey the standard post-Newtonian rules $\partial_\tau \sim \epsilon \partial_x \sim {\cal O}(\epsilon^4)$. Therefore, considering the leading order terms only, the EKG equations now read
\begin{eqnarray}
i\partial_\tau \psi &=& -\frac{1}{2x} \partial^2_x (x\psi) + U \psi \, ,  \label{e7} \\
\partial^2_x (xU) &=& x \psi \psi^\ast \, , \label{e8} \\
\partial_x U_2 &=& -x \psi^2 \, \label{e9}.
\end{eqnarray}
In addition, $A(\tau,x) = x \partial_x U$ and $A_2 \sim {\cal O}(\epsilon^4)$, that is, the metric coefficient $g_{rr}$ can be taken plainly as time-independent. 

That the complete system (\ref{e7}), (\ref{e8}) and (\ref{e9}) is really the weak-field manifestation of Eqs. (\ref{e6}) is easily verified since the former admits stationary solutions for $\psi = e^{-i\omega \tau}\phi(x)$. Under appropriate boundary conditions, such solutions are in turn the so-called Newtonian oscillatons \cite{luis}. Indeed, Eq.~(\ref{e9}) only arises for real scalar fields and represents the particular oscillatory behavior of the metric for oscillatons\cite{seidel91,miguel1,luis}.

It should be stressed here that the whole dynamics of the EKG system is contained in Eqs.~(\ref{e7}) and (\ref{e8}), which are the so-called Schr\"odinger--Newton (SN) equations \cite{lee,pang,arbey,seidel90,luis}; which also stands for the post-Newtonian expansion with complex scalar fields\cite{seidel90}.

Before proceeding further, we need to determine the initial scalar field profile. As we said before, prior to turnaround, the scalar profile is not determined by its own gravity only. In other words, the gravitational potential $U(\tau_{\rm turn},x)$ is not determined by Eq.~(\ref{e8})\cite{fuzzy}. But, $\psi$ should still obey the Schr\"odinger equation~(\ref{e7}), i.e., the initial condition should be in accord with the scalar nature of the matter fluctuation.  

Thus, the initial condition problem can be solved by finding the initial gravitational potential $U(0,r) = U(t_{\rm turn},r)$. The simplest assumption we can make is to approximate the gravitational potential by a spherical square well of the form \cite{seidel90}
\begin{eqnarray}
U(0,r < R_0) = U_0 \, , && \, \,  U(0,r > R_0) = 0 \, , \label{e10}
\end{eqnarray}
where $U_0=-GM_0/R_0$ is the depth of the gravitational well and $R_0$ is taken as the radius containing the total initial mass $M_0$. Next, we look for stationary solutions of Eq. (\ref{e7}) in the form $\psi=e^{-i\omega t} \phi(r)$, which is a common textbook problem on Quantum Mechanics\cite{quantum}. The initial scalar profile is then of the form (in dimensionless coordinates)
\begin{eqnarray}
\psi_i(0,x<x_0) &=& \psi_0 \frac{\sin (x/\sigma)}{(x/\sigma)} \, , \label{e15} \\
\psi_i(0,x>x_0) &=& \psi_1 \frac{\exp \left[-\sqrt{2|U_0|\sigma^2 - 1} \, (x/\sigma) \right]}{x} \, . \nonumber
\end{eqnarray}

The  eigenvalues of $\sigma$ are found by imposing the continuity of the radial function $\phi(r)$ and its first derivative at the discontinuity of the potential, and then are given by the solutions to the equation $\cot (x_0/\sigma) = - \sqrt{2|U_0|\sigma^2 - 1}$, with $x_0 = mR_0$. Hence, $\sigma$ can be estimated from the argument of the cotangent function as $(x_0/\sigma) \simeq (n+1)\pi < \sqrt{2|U_0|} x_0$, where $n=0,1,2,..$ labels the number of nodes of the initial profile. $\psi_1$ is evaluated from the continuity condition of the radial function at $x=x_0$. Finally, the initial mass according to (\ref{e15}) is
\begin{equation}
M_0 = 4\pi m^2 \int^\infty_0 \Phi^2 r^2 dr \simeq  \frac{1}{2} \frac{m^2_{\rm Pl}}{m} x_0 \sigma^2 \psi^2_0 \, , \label{e16}
\end{equation}
which becomes an equation that helps to determine $\psi_0$ for fixed $n,R_0,M_0$; a procedure rather convenient since the latter are the physical (input) parameters of the system.

We mention here an additional major simplification we can make use of for the SN equations, but {\it not} for the EKG ones. It is easily seen that Eqs. (\ref{e7}), (\ref{e8}) and (\ref{e9}) are invariant under the scaling transformation
\begin{equation}
\left\{ \tau,x,U,U_2,\psi \right\} \rightarrow \left\{ \lambda^{-2} \hat{\tau},\lambda^{-1} \hat{x}, \lambda^2 \hat{U}, \lambda^2 \hat{U}_2,\lambda^2 \hat{\psi} \right\} \label{e18}
\end{equation}
where $\lambda$ is an arbitrary parameter. A similar scaling property was noticed before for the stationary solutions of the SN equations, from which one realizes that all stationary solutions (for a fixed number of nodes of the field $\psi$) are related to each other by a scaling transformation \cite{pang,luis}. 

By means of~(\ref{e18}) and an appropriate $\lambda$, the collapse of our fluctuation can be reduced to the study of a conveniently sized configuration concerning {\it hat}-quantities only. Once the hat-configuration has been evolved, we apply the inverse transformation to recover the physical (no-hat) quantities. 

We focus now on the numerical solution to the SN equations. Once the initial profile $\psi_i(0,x)$ is given, the Poisson equation~ (\ref{e8}) is integrated with a second order accurate upwind method inwards under the boundary condition $U(\tau,x_f)=GM(\tau,x_f)/x_f$ (being $x_f$ the last point of our spherical grid), which is valid at each time slice according to the Newton theorems regarding spherical objects. The next scalar configuration is determined by solving the Schr\"odinger equation~(\ref{e7}) using a second order finite differences fully implicit Cranck-Nicholson method \cite{nr}. The procedure is then repeated forward in time.
 
To measure the accuracy of our solutions, we use the $\{ t,r \}$ component of the Einstein's equations, which rephrases the conservation of probability for the Newtonian field $\psi$,
\begin{equation}
\beta := \partial_t (\psi \psi^\ast) - \frac{i}{2x} \left[ \psi^\ast \partial^2_x (x\psi) - \psi \partial^2_x (x\psi^\ast) \right] \, , \label{constraint}
\end{equation}
where $\beta$ is zero for an exact solution. The accuracy depends on the time step $\Delta \tau$ and the grid resolution $\Delta x$, which should be chosen appropriately to assure that $|\Delta \psi|/|\psi| \ll 1$ in a time step. Thus, we should comply with both $\Delta \tau/ (\Delta x)^2 \leq 1$ and $[\Delta \tau/ (\Delta x)^2] |1+U(\Delta x)^2| < 1$. The former is the condition applied to a free wave function, and the latter takes into account the presence of a potential in Eq.~(\ref{e7}).

\begin{figure}[t]
\includegraphics[width=4cm]{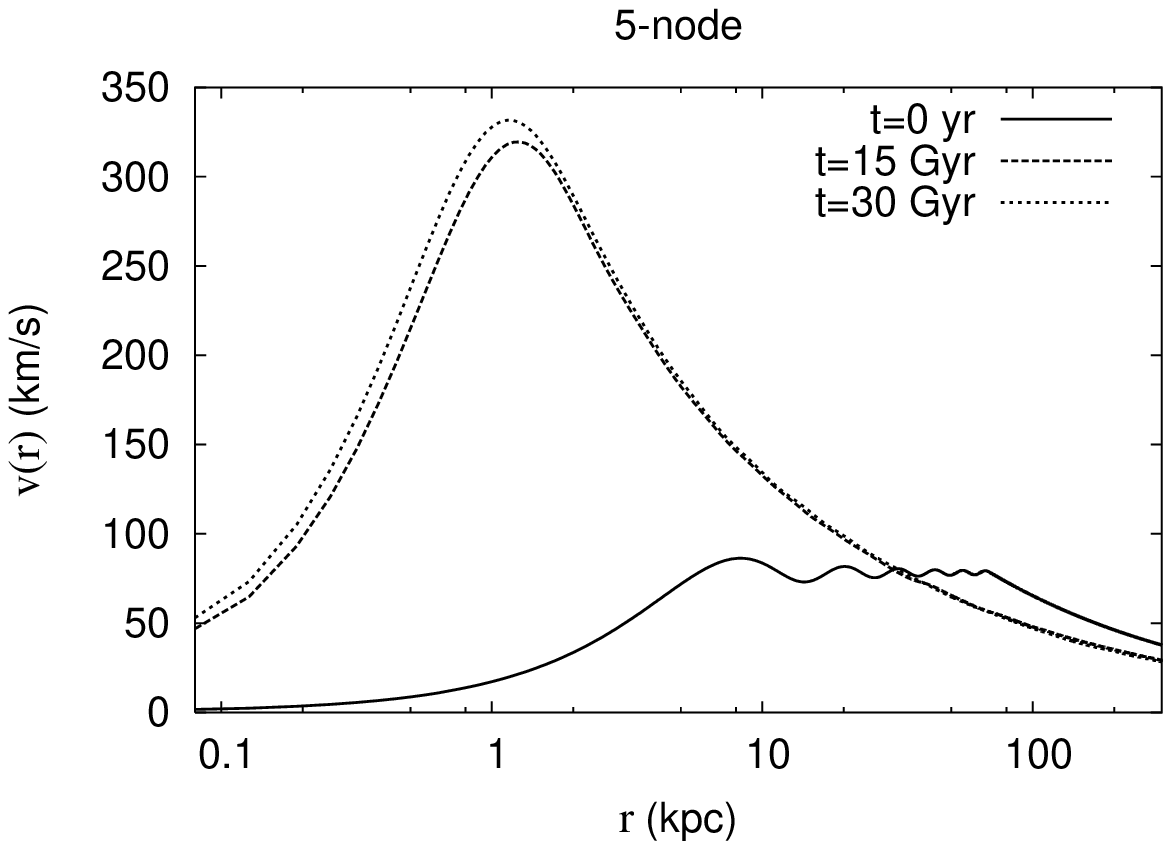}
\includegraphics[width=4cm]{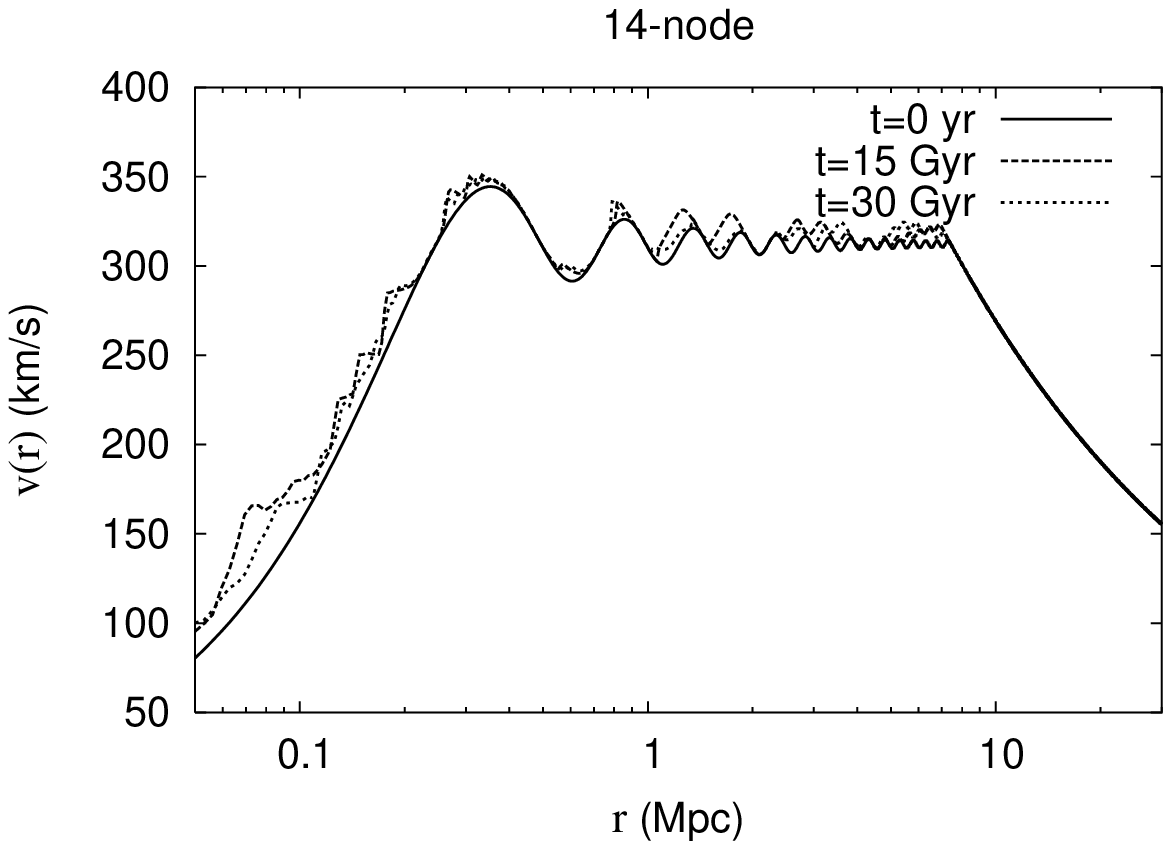}
\includegraphics[width=4cm]{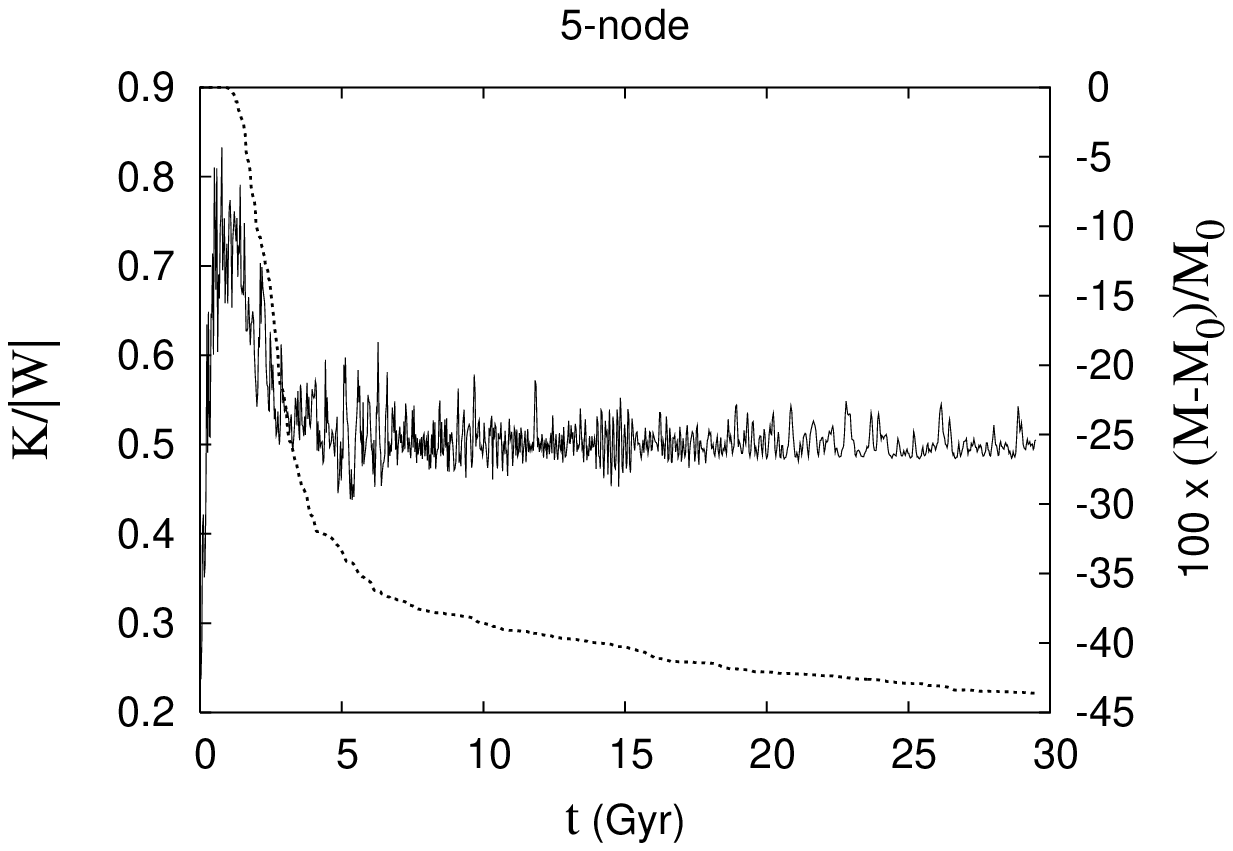}
\includegraphics[width=4cm]{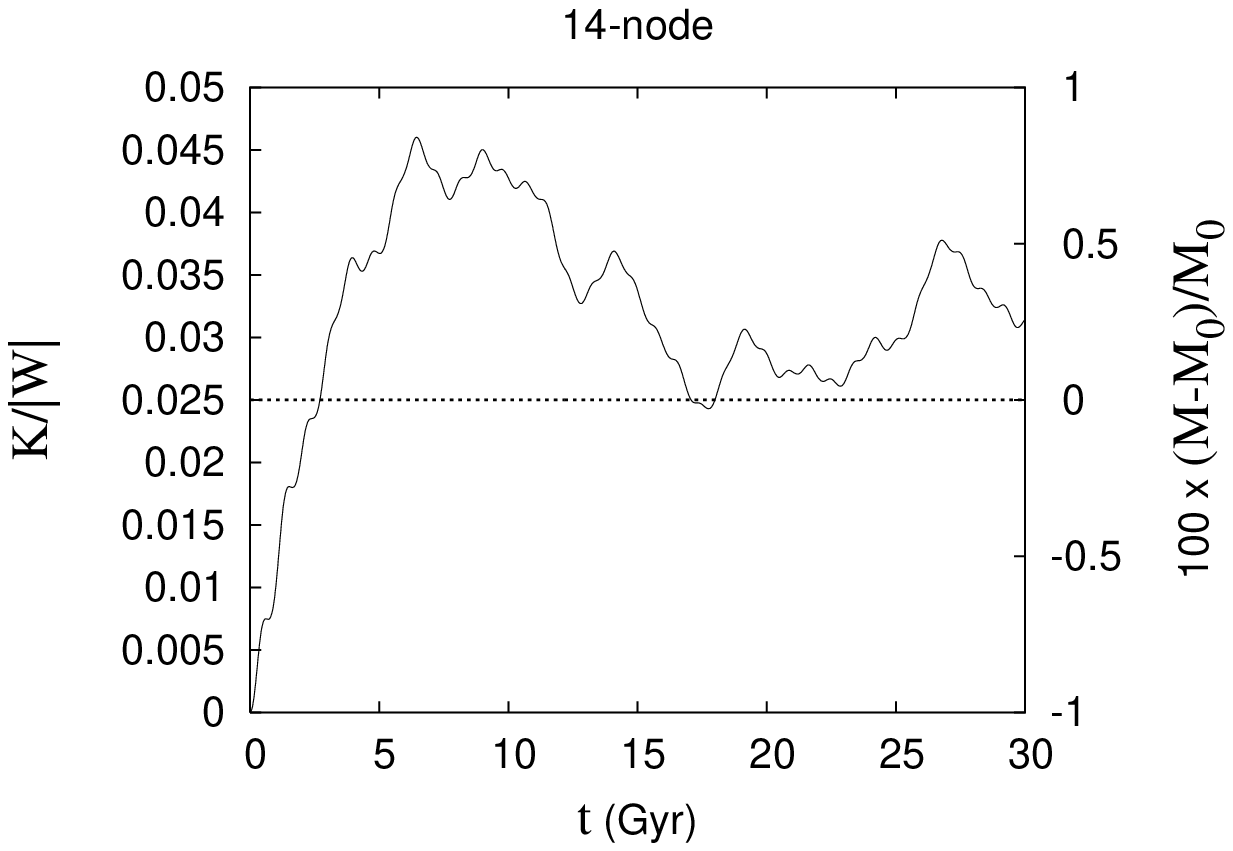}
\caption{\label{fig:f1} Evolution of a 5-node and a 14-node initial scalar fluctuation of the form (\ref{e15}), see the text for details. Top: Profiles of the rotation curves $v(r) = \sqrt{{\rm G}M(r)/r}$ at different times. Bottom: Evolution of the ratio ${\rm Kinetic \, Energy/|Gravitational \, Potential \, Energy|}$ (solid line), which shows the virialization of the two systems, and the total integrated mass $M$ (dotted line). It is worth to notice the so called gravitational cooling after free fall for the first of the systems, which is not achieved in the second case within the time scale shown, but is expected afterwards.}
\end{figure}

In Fig.~\ref{fig:f1} we show the runs of an initial $5$ $(14)$-node scalar fluctuation for a single scalar halo, whose initial physical parameters are $M_0=10^{11}~(1.7 \times 10^{14})~M_\odot$ and $R_0=70~{\rm kpc}~(7.3~{\rm Mpc})$. The corresponding scale parameter is $\lambda^2=6.38~(0.6) \times 10^{-6}$, and then $\hat{\sigma}=14.6~(190)$. The grid spacing is $\Delta \hat{x}=0.25~(4.0)$ with the boundary at $\hat{x}_f=1250~(4\times 10^{4})$. The time step is $\Delta \hat{\tau}= 3~(0.125) \times 10^{-2}$, and the runs were followed up to a physical time of $T_0 = 30 \, {\rm Gyr}$. In all cases, $\| \beta \|_{2} < 10^{-7}$. 

We see in Fig.~\ref{fig:f1} that the initial rotation curve profile of both evolved configurations is almost flat, but such flatness is lost at late times during the collapse. The smallest configuration, which is supposed to have started collapsing earlier within the cosmological picture, rapidly virializes and forms a giant Newtonian boson star, though not as large as a realistic (dwarf) galaxy. According to Eq.~(\ref{e18}), the same fate is expected for the largest configuration, which roughly corresponds to a cluster of galaxies fluctuation, but within a time scale much larger than the actual age of the universe.

Some final comments are in turn. The first one is about the initial profile~(\ref{e15}). It is well known that the self-gravitating equilibrium configuration of the SN equations (the so-called Newtonian boson stars\cite{pang} and Newtonian oscillatons \cite{luis}) are reached by the system after the scalar field has had enough time to modify the gravitational potential. We assume that such condition is not achieved before turnaround because there are other factors involved in determining the scalar profile. Therefore, one does not expect the initial profile to be a Newtonian oscillaton, but the latter will rather be the {\it final} configuration of the evolutions shown here. Indeed, Fig.~\ref{fig:f1} confirms the expectations and shows how, in some cases, a scalar field fluctuation can virialize and form a Newtonian oscillaton. Second, we have seen that a model of structure formation within the scalar field hypothesis could imply an initial configuration with nodes. The latter can be naturally expected since the Compton length of the scalar particles is, typically, much smaller than the size of the initial configuration, $\lambda_C = m^{-1} \ll R_0$. Then, the wave function can be confined and {\it wrinkled} by the gravitational well potential (which remains fixed in time, see \cite{liddle,coles,bagla}) before the non-linear collapse of the fluctuation. 

On the other hand, the scale invariance of the SN equations~(\ref{e18}) suggests that all galaxies should look similar; in particular, some of them could be related just through a scaling transformation. This argument can be extended to larger and more massive fluctuations, and then it predicts that the latter should have similar properties as galaxies.

The SN equations were derived under the assumption of spherical symmetry, but we really expect the EKG equations to have a tractable non-spherical weak-limit too. This would suggest that the full SN equations without symmetries should be used to explore the realization of a realistic non-spherical collapse and to track the collapse of many scalar fluctuations. Also, the arguments about the initial configurations which lead us to~(\ref{e15}) can be extended to the cases of non-zero angular momentum (see for instance \cite{widrow,coles}).

Another advantage of the SN system is that the scalar field galaxy halo can be, in an easy manner, gravitationally coupled to baryons; the latter would only enter on the r.h.s. of the Poisson equation~(\ref{e8}) \cite{arbey}. Results on this issue and others, like more realistic initial conditions and the study of rotation curves profiles, will be reported elsewhere.

Finally, all the results shown here are also valid for complex scalar fields, since the dynamics of both type of scalar fields is determined by the SN system only, though real scalar fields would be preferred from the cosmological point of view \cite{peebles,arbey}.

Summarizing, we have shown how to track the non-linear collapse of a matter fluctuation made of real scalar fields using the Schr\"odinger-Newton system, and the kind of properties it should have before and after separating from the cosmological expansion.

%======================================%
%<<<<<<<<<<< ACKNOWLEDGEMENTS >>>>>>>>>>%
%======================================%

\begin{acknowledgments}
We want to thank Miguel Alcubierre, Ricardo Becerril, Andrew Liddle, Tonatiuh Matos, Dar\'{\i}o N\'u\~nez and Ed Seidel for a lot of important insight, Erasmo G\'omez and Aurelio Esp\'{\i}ritu for technical support. The runs were carried out on machines of the Laboratorio de Superc\'omputo Astrof\'{\i}sico at CINVESTAV. This research was partly supported by CONACyT, M\'exico under grant 010385 (L.A.U.-L.), and by a bilateral project DFG-CONACyT 444 MEX-13/17/0-1.
\end{acknowledgments}

%======================================%
%<<<<<<<<<<<< BIBLIOGRAPHY >>>>>>>>>>>>%
%======================================%

%%%%%%%%%%%%%%%%%%%%%%%%%%%%%%%%%%%%%%%%%%%%%%%%%%%%%%%%%%%%%%%%%%%%%%%%

\begin{thebibliography}{}
\bibitem{max} D. N. Spergel {\it et al}, {\tt astro-ph/0302209}. J. L. Sievers {\it et al}, {\tt astro-ph/0205387}.
\bibitem{primack} J. R. Primack, {\tt astro-ph/0112255}, {\tt astro-ph/0205391}.
\bibitem{prada} F. Prada {\it et al}, {\tt astro-ph/0301360}.
\bibitem{dekel} A. Dekel, J. Devor, and G. Hetzroni, {\tt astro-ph/0204452}.
\bibitem{sin} S. J. Sin, Phys. Rev. D {\bf 50}, 3650 (1994); S. U. Ji and S. J. Sin, Phys. Rev. D {\bf 50}, 
	3655 (1994).
\bibitem{lee} J. W. Lee and I. G. Koh, Phys. Rev. D {\bf 53}, 2236 (1996).
\bibitem{peebles} P. J. E. Peebles, Astrophys. J. {\bf 534}, L127 (2000);
	J. Goodman, New Astron. {\bf 5}, 103 (2000).
	M. C. Bento, O. Bertolami, R. Rosenfeld, and L. Teodoro, Phys.
	Rev. D {\bf 62}, 041302 (2000); O. Bertolami, 
	M. C. Bento, and R. Rosenfeld, astro-ph/0111415. A. Riotto and I. Tkachev, Phys. Lett. B {\bf 484}, 177 (2000);
	V. Sahni and L. Wang, Phys. Rev. D {\bf 62}, 103517 (2000);
	T. Matos and L. A. Ure\~na-L\'opez, Class. Quantum Grav.
	{\bf 17}, L75 (2000), Phys. Rev. D {\bf 63},
	063506 (2001), Phys. Lett. B {\bf 538}, 246
	(2002); J. E. Lidsey, T. Matos, and L. A.
	Ure\~na-L\'opez, Phys. Rev. D {\bf 66}, 023514 (2002); 
        L. A. Ure\~na-L\'opez and A. R. Liddle, Phys. Rev. D {\bf 66}, 083005 (2002).
\bibitem{arbey} A. Arbey, J. Lesgourgues, and P. Salati, Phys. Rev. D {\bf 64},
	123528 (2001); {\it ibid}, {\bf 65}, 083514
	(2002); {\tt astro-ph/0301533}.
\bibitem{miguel} M. Alcubierre, F. S. Guzm\'an, T. Matos, D. N\'u\~nez, 
	L. A. Ure\~na-L\'opez, and P. Wiederhold, Class. Quantum Grav. {\bf 19}, 5017 (2002).
\bibitem{fuzzy} W. Hu, R. Barkana, and A. Gruzinov, Phys. Rev. Lett. {\bf 85}, 1158         (2000). 
\bibitem{mielke}E. W. Mielke and F. E. Schunck, Phys. Rev. D {\bf 66}, 023503 (2002); 
        R. P. Yu and M. J. Morgan, Class. Quantum Grav. {\bf 19}, L157(2002).
\bibitem{seidel90} E. Seidel and W-M. Suen, Phys. Rev. D {\bf 42}, 384 (1990);
        J. Balakrishna, E. Seidel, and W-M. Suen, Phys. Rev. D {\bf 58}, 104004 (1998).
\bibitem{seidel91} E. Seidel and W-M. Suen, Phys. Rev. Lett. {\bf 66}, 1659
	(1991). E. Seidel and W-M. Suen, Phys. Rev. Lett. {\bf 72}, 2516
	(1994).
\bibitem{miguel1} M. Alcubierre, R. Becerril, F. S. Guzm\'an, T. Matos, D. N\'u\~nez, and L. A. Ure\~na-L\'opez, gr-qc/0301105.
\bibitem{liddle} A. R. Liddle and D. H. Lyth, {\it Cosmological inflation and          large-scale structure} (Cambridge University Press, 2000). T. Padmanabhan, {\it Structure formation in the universe} (Cambridge University Press, 1993). 
\bibitem{luis} L. A. Ure\~na-L\'opez, Class. Quantum Grav. {\bf 19}, 2617
	(2002); L. A. Ure\~na-L\'opez, T. Matos, and 
	R. Becerril, Class. Quantum Grav. {\bf 19}, 6259 (2002).
\bibitem{pang} R. Friedberg, T. D. Lee, and Y. Pang, Phys. Rev. D {\bf 35}, 
	3640 (1987). I. M. Moroz, R. Penrose, and P. Tod, Class. Quantum Grav.
	{\bf 15}, 2733 (1998); P. Tod and I. M. Moroz, Nonlinearity {\bf 12},
	201 (1999).
\bibitem{quantum} E. Merzbacher, {\it Quantum Mechanics} (John Wiley \& Sons,               1998).
\bibitem{nr} W. H. Press, S. A. Teukolsky, W. T. Watterling and B. P. Flannery, 
        {\it Numerical Recipes in Fortran}. Cambridge University Press, 1992.
\bibitem{coles} P. Coles and K. Spencer, {\tt astro-ph/0212433}.
\bibitem{bagla}J. S. Bagla and T. Padmanabhan, {\tt gr-qc/9304021}.
\bibitem{widrow} G. Davies and L. M. Widrow, Astrophys. J. {\bf 485}, 484 (1997).

\end{thebibliography}
\end{document}